\documentstyle[prl,aps]{revtex}
\begin{document}
\draft
\title{Quantum-classical correspondence and nonclassical states generation
in dissipative quantum optical systems}
\author{Kirill N. Alekseev$^{1,2}$\cite{email1}, Natasha V. Alekseeva$^{1}$,
and Jan Pe\v{r}ina$^{2}$\cite{email2}}
\address{$^1$Kirensky Institute of Physics,
Russian Academy of Sciences, Krasnoyarsk 660036, Russia\\
$^2$Department of Optics and Joint Laboratory of Optics\\
 Palack\'{y} University, 17. listopadu 50, 772 07 Olomouc, Czech Republic}
\maketitle
\begin{abstract}
We develop a semiclassical method for the determination of the nonlinear
dynamics of dissipative quantum optical systems in the limit of large number
of photons $N$, based on the $1/N$-expansion and the quantum-classical
correspondence. The method has been used to tackle two problems: to study
the dynamics of nonclassical state generation in higher-order anharmonic
dissipative oscillators and to establish the difference between the quantum
and classical dynamics of the second-harmonic generation in a self-pulsing
regime. In addressing the first problem, we have obtained an explicit time
dependence of the squeezing and the Fano factor for an arbitrary degree of
anharmonism in the short-time approximation. For the second problem, we have
established analytically a characteristic time scale when the quantum dynamics
differs insignificantly from the classical one.
\end{abstract}
\pacs{05.45.+b, 03.65.Sq, 42.50.Dv}
\section{Introduction}
\label{sec:introduc}
The situation when nonlinear interactions involve a large number of photons,
$N$, is quite typical of many of the problems of quantum and nonlinear optics
\cite{1,Fabre,Reynaud}. Hiedmann {\it et al.} suggested \cite{7} to use the
method of the $1/N$-expansion \cite{cumulant} to describe the nonlinear
dynamics of the mean values and second-order cumulants of a quantum system
in the limit $N\gg 1$. Following the general scheme of that method
\cite{cumulant}, an exact or approximate solution can be found first in terms
of the coherent state representation in the classical limit
$N\rightarrow\infty$ and then adjusted by adding the quantum corrections.
The method proves to be particularly convenient when the generation
dynamics of nonclassical states needs to be determined \cite{7}. We have
recently developed the method further to study the enhanced squeezing at
the transition to quantum chaos \cite{8,9,9r}.
\par
Papers \cite{7,8,9} are concerned with consideration of the problems of
nondissipative quantum systems only. In this paper we extend the method to
dissipative quantum systems. For quantum systems without dissipation, the
lowest-order of the $1/N$-expansion is equivalent to the linearization in
terms of the classical solution \cite{8,9}, whereas in dissipative systems,
as will be demonstrated herein, the solution of motion equations for
variations near the classical trajectory cannot provide complete information
on the dynamics of quantum fluctuations even in the lowest order of $1/N$.
We will show that the influence of reservoir on the dynamics of
the expectation values and dispersions, which is different from the energy
dissipation,
always exists; it has the quantum nature and can not be neglected even in the
semiclassical limit. Specific manifestations of the effect will however
depend on the type of the attractor in the underlying classical dynamic system.
For systems with a simple attractor in the classical limit, the ``quantum
diffusion'' associated with the reservoir quantum fluctuations will not
introduce any new physical effects in the dynamics of the main system, at
least in the short-time limit. In contrast to that, for a stable limit cycle,
such a diffusion appears to be the main mechanism responsible for the
difference between the classical and quantum dynamics at $N\gg 1$.
\par
Along with the presentation of a general formalism, two typical examples of
quantum optical systems with a simple attractor and a stable limit cycle
in the classical limit $N\rightarrow\infty$ will be considered: the dissipative
higher-order anharmonic oscillator and the self-pulsing regime of
intracavity second-harmonic generation (SHG). We will show how the method of
$1/N$-expansion can be used to investigate the dynamics of nonclassical state
generation and to determine a time scale for a correct classical description
of the dissipative quantum dynamics.
\par
A quantum anharmonic oscillator with the Kerr-type nonlinearity is one of
the simplest and most popular models employed in the description of quantum
statistical properties of the light interacting with a nonlinear medium
\cite{1,2}. The Kerr oscillator model with a third-order nonlinearity yields
an exact solution in both nondissipative \cite{4} and dissipative limits
\cite{2}. However, due to the complexity of the solution in the dissipative
case, numerical methods or special approximate analytical methods have to be
used to determine statistical properties of the radiation in a most relevant
experimental case involving a large number of photons. Moreover, there are no
exact solutions available for the model of the anharmonic oscillator with a
higher-order nonlinearity.
\par
In this paper, a simple and explicit time dependencies of the degree of
squeezing and the Fano factor are obtained analytically in the model of
anharmonic oscillator of an arbitrary order for the most interesting
experimental situation featuring higher intensities ($N\gg 1$) and
short-time interactions.
As another example of application of the $1/N$-expansion, we consider the
self-pulsing in SHG \cite{drummond}. Such an oscillatory regime, corresponding
to the limit cycle, was observed experimentally in \cite{kimble-exp}. There are
quite a few papers dealing with the development of approximate analytical and
numerical methods with the purpose of describing different dynamic
regimes in SHG in terms of quantum mechanics
\cite{perina-cz-j,savage-pra,schack,szlachetka,savage-comp}.
In particular, Savage \cite{savage-pra} calculated the $Q$ distribution
function in the Gaussian approximation about the classical limit cycle. He
demonstrated numerically that in the classical limit, the initial rapid
collapse of the Q distribution in the neighbourhood of the limit cycle was
followed by the diffusion around the limit cycle. However, the author did not
offer any analytical solution of the problem or explanation of the
physics of the effect observed.
\par
In this paper we show that the diffusion around the classical limit cycle can
be obtained as a solution  of the motion equations for low-order cumulants by
using the $1/N$-expansion technique.
This enables us to find the time scale $t\ll t^*$ with $t^*\simeq 2 N
\gamma^{-1}$ ($\gamma$ is a damping constant), for a correct classical
description of self-oscillations in SHG. The resultant estimate is consistent
with that obtained for $t^*$ numerically in \cite{savage-pra}. Finally, we
interpret the quantum diffusion around the limit cycle as having been caused
by the effect of the reservoir vacuum on the SHG dynamics.
\par
The structure of the paper is as follows. Section \ref{sec:1/N-expansion}
describes a general formalism of the $1/N$-expansion applicable to an
arbitrary single-mode quantum dissipative system and presents the
solution of the motion equations for mean values and second-order cumulants
obtained in the first order of $1/N$. Sections \ref{sec:Noncl_stats} and
\ref{sec:SHG} deal with the nonclassical state generation dynamics in
higher-order anharmonic oscillators and the quantum-classical correspondence
for the self-pulsing regime in SHG, respectively. The final section contains
a summary and concluding remarks.

\section{$1/N$-expansion and quantum-classical correspondence}
\label{sec:1/N-expansion}

First of all, we need to generalize the approach of \cite{9} for the case
of systems with dissipation. As an illustrative example we consider a quantum
anharmonic oscillator with the Hamiltonian in the interaction
picture
\begin{equation}\label{1} H=\Delta b^{\dag}b + \frac{\lambda_l}{l+1}
\left( b^{\dag} b \right)^{l+1},\quad [ b, b^{\dag} ]=1,\end{equation}
where the operators $b$ and  $b^{\dag}$ describe a single mode of a quantum
field and the constant $\lambda_l$ is proportional to a $(2 l+1)$-order
nonlinear susceptibility of a nonlinear medium ($l$ is an integer), $\Delta$ is
the light frequency detuning from the characteristic frequency of quantum
transition, and $\hbar\equiv 1$. Everywhere in this paper we use the
normal ordering of operators. The oscillator interacts with an infinite
linear reservoir of finite temperature. The Hamiltonians of the reservoir and
interaction of the oscillator with reservoir are defined as follows
\begin{equation}
\label{2}
H_{\rm r}=\sum_{j} \psi_j (d_j^{\dag}d_j+1/2), \quad
H_{\rm int}=\sum_{j}\left(\kappa_j d_j b^{\dag} + {\rm H. c.}\right),
\end{equation}
where the Bose operator  $d_j$ ($[d_j, d_k^{\dag}]=\delta_{jk}$) describes
an infinite reservoir with characteristic frequencies $\psi_j$, and
$\kappa_j$ are the coupling constants between reservoir modes and
the oscillator.
Introduce new scaled operators $a=b/N^{1/2}$, $c_j=d_j/N^{1/2}$
and the Hermitian conjugates satisfying the commutation relations
\begin{equation} \label{4}
[ a, a^{\dag} ]=1/N, \quad [c_j, c_k^{\dag}]=\delta_{jk}/N.\end{equation}
In the classical limit $N\rightarrow\infty$, we have commuting classical
$c$-numbers instead of operators. Now the full Hamiltonian
$H=H_0+H_r+H_{int}$ may be rewritten as $H=N {\cal H}$, where ${\cal H}$ has
the same form as (\ref{1}) and (\ref{2}) but for the following
replacements
\begin{equation}\label{5}
b\rightarrow a,\quad b^{\dag}\rightarrow a^{\dag},\quad
d_j\rightarrow c_j,\quad d_j^{\dag}\rightarrow c_j^{\dag},\quad
\mbox{and}\quad \lambda_l\rightarrow g_l(N)\equiv \lambda N^l.\end{equation}
It can be shown that the photon-number dependent constant $g_l(N)$
provides a correct time scale of oscillations for the nonlinear
oscillator (\ref{1}) in the classical limit (for the case of Kerr
nonlinearity with $l=1$, see, {\it e. g.} \cite{10}).
Note that ${\cal H}$ can have an explicit time dependence in the
general case \cite{9}. Within a standard
Heisenberg-Langevin approach, the equation of motion has the form
(\cite{1}, chap. 7)
\begin{equation}\label{5'}
\dot{a}=-i\left(\Delta-i\frac{\gamma}{2}\right) a + V +L(t),\end{equation}
where $V=\partial{{\cal H}_0}/\partial a^{\dag}$,
$\gamma=2\pi|\kappa(\omega)|^2\rho(\omega)$ is the damping constant,
$\rho(\omega)$ being the density function of reservoir oscillators,
which spectrum is considered to be flat. The Langevin force operator $L(t)$ is
in a standard relation to the operators $\{c_j\}$ of the reservoir \cite{1}.
The properties of $L(t)$ \cite{1} in our notations (\ref{5}) may be rewritten
as
\begin{equation}\label{6}
\langle L(t)\rangle_R=\langle L^{\dag}(t)\rangle_R=0,\quad
\langle L^{\dag}a\rangle_R+\langle a^{\dag}  L\rangle_R=
\gamma\frac{\langle n_d\rangle}{N},\quad
\langle L a\rangle_R+\langle a L\rangle_R=0,\end{equation}
where the averaging is performed over the reservoir variables and
$\langle n_d\rangle$ is a single-mode mean number of reservoir quanta
(phonons), that is related to temperature $T$ as
$\langle n_d\rangle=\left[\exp\left(\frac{\omega}{k T}\right) - 1
\right]^{-1}$, where $k$ is the Boltzmann constant and $\omega$ is the
characteristic phonon frequency. From the Heisenberg-Langevin equations for
$a$, $a^2$ and the Hermitian conjugated equations, by using Eqs. (\ref{5'})
and (\ref{6}), we obtain:
\begin{eqnarray}
\label{7}
i \frac{d}{dt}\langle\alpha\rangle & =&
\langle V\rangle-i\frac{\gamma}{2}\langle\alpha\rangle, \nonumber\\
i \frac{d}{dt}\langle\left(\delta\alpha\right)^2\rangle &= &
2\langle V\delta\alpha\rangle+
\langle W\rangle
-i\gamma\langle\left(\delta\alpha\right)^2\rangle, \\
i \frac{d}{dt}\langle\delta\alpha^*\delta\alpha\rangle & = &
-\langle V^*\delta\alpha\rangle+
\langle\delta\alpha^* V\rangle
-i\gamma\langle\delta\alpha^*\delta\alpha\rangle
+i\gamma\frac{\langle n_d\rangle}{N}, \nonumber
\end{eqnarray}
where $W=(1/N)\partial V/ \partial a^{\dag}$,
$z\equiv\langle a\rangle$,
$\langle\left(\delta\alpha\right)^2\rangle=\langle a^2\rangle-z^2$,
$\langle\delta\alpha^*\delta\alpha\rangle =
\langle a^{\dag} a\rangle-|z|^2$, and the averaging is performed over
both the reservoir variables and the coherent state
$|\alpha\rangle=\exp(N\alpha a^{\dag} -N\alpha^* a) |0\rangle$
corresponding to the mean photon number $\simeq N$. In deriving Eq.
(\ref{7}), we neglect the insignificant additional detuning introduced to
$\Delta$ by the interaction with the reservoir \cite{1}. In the absence of no
damping $\gamma=0$, our equations for the mean values and the second-order
cumulants (\ref{7}) are reduced to the corresponding equations in \cite{7,9}.
\par
Set of equations (\ref{7}) is not closed and is basically equivalent
to the infinite  dynamical hierarchy system for the cumulants of a different
order. To truncate it up to the second-order cumulants, we make the
substitution $a\rightarrow z +\delta\alpha$,
where at least initially the mean $z\simeq1$ and
the quantum correction $|\delta\alpha(t=0)|\simeq N^{-1/2}\ll 1$.
Using the Taylor expansion of the functions $V$ and $W$
and after some algebra analogous to that used in \cite{9},
we get  from (\ref{7}) in the first order of $1/N$ the following
self-consistent system of equations for the mean value and the second order
cumulants (for details see \cite{preprint})
\begin{mathletters} \label{9}
\begin{equation}\label{9a}
i\dot{z}=-i\frac{\gamma}{2} z+
\langle V\rangle_z + \frac{1}{N} Q(z, z^*, C, C^*, B),
\end{equation}
\begin{equation}\label{9b}
i\dot{C}=2\left(\frac{\partial V}{\partial\alpha}\right)_z C +
2\left(\frac{\partial V}{\partial\alpha^*}\right)_z B-i \gamma C,
\end{equation}
\begin{equation}\label{9c}
i\dot{B}=-\left(\frac{\partial V^*}{\partial\alpha}\right)_z C +
\left(\frac{\partial V}{\partial\alpha^*}\right)_z C^*
-i\gamma\left( B-B^{(0)}\right)
\end{equation}\end{mathletters}
and the corresponding equation for $C^*(t)$ that could be obtained from
equation (\ref{9b}) by way of complex conjugation. The quantum correction
to the classical motion $Q$ in Eq. (\ref{9a}) has the following form
\begin{equation}\label{10}
Q=
\frac{1}{2} \left(\frac{\partial^2 V}{\partial\alpha^2}\right)_z C +
\frac{1}{2} \left(\frac{\partial^2 V}{\partial\alpha^{*2}}\right)_z C^* +
\left(\frac{\partial^2 V}{\partial\alpha^{*}\partial\alpha}\right)_z
\left(B-\frac{1}{2}\right).\end{equation}
In Eqs. (\ref{9}) and (\ref{10}) the
subscript $z$ means that the values of $V$ and its derivatives
are calculated for the mean value $z$ and we have introduced
\begin{equation}\label{B_and_C}
B=N\langle\delta\alpha^*\delta\alpha\rangle+1/2,\quad
C=N\langle\left(\delta\alpha\right)^2\rangle.
\end{equation}
The initial conditions for system (\ref{9}) are
\begin{equation}\label{initial_cond}
B(0)=1/2,\quad C(0)=0,\end{equation}
and an arbitrary $z(0)\equiv z_0$ which is of order unity.
The equilibrium value of cumulant $B$ in equation (\ref{9c}) is determined
by the mean number of reservoir's quanta and its zero-point energy as
\begin{equation}\label{B_0}
B^{(0)}=\langle n_d\rangle+1/2.\end{equation}
Note that the zero-point energy of the reservoir appears in the
equations of motion
for the cumulants though it was not presented in the Heisenberg equations
of motion and even may be dropped from the Hamiltonian
redefining a zero of energy. Such ``reappearance''
of a zero-point field energy is quite common in other problems of quantum
theory where a vacuum is responsible for the physical effects
\cite{milonni_book}.
\par
The motion equations for second-order cumulants $B$ and $C$ [Eqs. (\ref{9b}),
(\ref{9c})] are
linear inhomogeneous equations. Their solution consists of two parts: a
general solution of the homogeneous set of equations
({\it i.e.} without term $+i\gamma B^{(0)}$ in Eq. (\ref{9c})) that we
denote as $\left( \overline{B}(t),\overline{C}(t)\right)$,
and the particular solution of the inhomogeneous equations
\begin{equation}\label{solution_struc}
\left(B(t),C(t)\right)=\left( \overline{B}(t),\overline{C}(t)\right)+
\left( \gamma B^{(0)} t, 0\right).
\end{equation}
To find $\left( \overline{B}(t),\overline{C}(t)\right)$ we use the
perturbation theory for $N\gg 1$ and as a first step neglect the quantum
correction $Q/N$ in Eq. (\ref{9a}). It is easy
to see that the homogeneous equations of motion for cumulants (\ref{9b})
and (\ref{9c})
can be obtained from the classical equation ({\it i.e.} from (\ref{9a}) with
$Q/N\rightarrow 0$) by linearization
around $z$ (substitution $z\rightarrow z+\delta z$, $|\delta z|
\ll |z|$), if one writes the dynamical equations for the variables
$(\delta z)^2$ and $|\delta z|^2$. The only difference between the
linearization of classical motion equations and equations for
quantum cumulants (\ref{9b}), (\ref{9c}) lies in the impossibility
to get the initial conditions (\ref{initial_cond}) for $C$ and $B$
from only initial conditions
for linearized classical equations of motion (see also the discussion of
this problem in \cite{9}). Hence, we first need to know the classical solution
$z_{\rm cl}(t)$, find differentials $dz_{\rm cl}$ and $dz_{\rm cl}^*$, and
then use the substitution
$\left( \overline{B}(t),\overline{C}(t)\right)\rightarrow
\left( |dz|^2 , (dz)^2 \right)$.
\par
Thus, it has become apparent that assuming the actual field deviates little
from the coherent state and treating the small deviation as
as a first-order correction would not be equivalent to direct linearization
around a classical trajectory.
Even in the limit $N\rightarrow\infty$, we will always deal with the
influence of reservoir on the dynamics of the quantum system via
the second-order cumulant $B$, which has the form of quantum diffusion
\begin{equation}\label{quant_dif}
B(t)=\overline{B}(t)+(\langle n_d\rangle+1/2)\gamma t,\end{equation}
where $\overline{B}$ is obtained from linearization around a large mean field.
In particular, as follows from Eq. (\ref{quant_dif}), the quantum diffusion
also exists for the case of a quiet reservoir $\langle n_d\rangle=0$.
\par
We now discuss the range of validity of the $1/N$-expansion and the role of
quantum diffusion in different classical dynamical regimes.
The criterion of validity of the $1/N$-expansion may be represented in
two forms. First, the $1/N$-expansion works well, provided the difference
between the classical and quantum solutions is small
\begin{equation}\label{Q-criterion}
\left|\frac{z(t)-z_{\rm cl}(t)}{z_{\rm cl}(t)}\right|\simeq
\frac{1}{N}\frac{\left|\int^t Q(t') dt'\right|}{|z(t)|}\ll 1,
\end{equation}
where $z_{\rm cl}(t)$ is the solution of Eq. (\ref{9a}) for
$N\rightarrow\infty$.
To write the second form of the criterion of validity of the $1/N$-expansion
we introduce following \cite{8,9} the ``convergence radius''
$R=\left\{ [ {\rm Re}(\delta\alpha) ]^2+[ {\rm Im}(\delta\alpha)]^2
\right\}^{1/2}$. Then, the expansion is correct within a time interval when
\begin{equation}\label{R-criterion}
\frac{R(t)}{|z(t)|}\simeq\frac{B^{1/2}(t)}{N^{1/2}|z(t)|}\ll 1.
\end{equation}
As a rule, the both conditions, Eqs. (\ref{Q-criterion}) and (\ref{R-criterion}),
determine the same time interval for validity of the $1/N$-expansion \cite{8,9}
(For a physically interesting exception, the problem of SHG, see Sec.
\ref{sec:SHG}).
\par
For dissipative systems with a simple attractor, the classical field intensity
$|z_{\rm cl}(t)|^2$, as well as cumulants $\overline{B}(t)$, $C(t)$ and
quantum correction $Q(t)$ are proportional to the factor $\exp(-\gamma t)$
and therefore, as follows from Eqs. (\ref{Q-criterion}) and
(\ref{R-criterion}) with account of Eq. (\ref{quant_dif}),
the $1/N$-expansion is well defined only in the time
interval of order of several relaxation times: $t^*\simeq\gamma^{-1}$
\cite{preprint}.
Moreover, during this time interval the influence of quantum diffusion
on the system dynamics is small.
\par
A quite different behavior is characteristic for the stable limit cycle. Here,
a variation near classical trajectory collapses to zero
($\delta\alpha\rightarrow 0$) and
therefore $\overline{B}(t)\simeq |\delta\alpha|^2\rightarrow 0$,
$C(t)\simeq (\delta\alpha)^2\rightarrow 0$. However $|z_{\rm cl}(t)|\simeq 1$
for the limit cycle and, as a result, the time interval of validity of the
$1/N$-expansion is fairly large $t^*\simeq N\gamma^{-1}$. What is important,
the diffusion is the major physical mechanism responsible for the difference
between the classical and quantum dynamics for a stable limit cycle.
In the two following sections we consider two typical examples
of dissipative optical systems with a simple attractor and limit cycle.

\section{Nonclassical states generation in higher-order anharmonic
oscillators}
\label{sec:Noncl_stats}

We start by defining the squeezing and the Fano factor.
Define the general field quadrature as $X_\theta=a\exp(-i\theta)+
a^{\dag}\exp(i\theta)$, where $\theta$ is the local oscillator phase.
A state is said to be squeezed if there exists some value $\theta$
for which the variance of $X_\theta$ is smaller than the variance for a
coherent state or the vacuum \cite{1,2}. Minimizing the variance of
$X_\theta$ over $\theta$, we get the condition of so-called principal squeezing
\cite{1,2,4} in the form
\begin{equation} \label{11}
S\equiv 1+2 N (\langle|\delta\alpha|^2\rangle-
|\langle(\delta\alpha)^2\rangle|)=2 (B-|C|) < 1.\end{equation}
The determination of the principal squeezing $S$ is very useful because it
gives the maximal squeezing measurable by the homodyne detection \cite{1,2}.
\par
Another important characteristic of nonclassical properties of the
light is the Fano factor
$F=(\langle n^2\rangle-\langle n\rangle^2)/\langle n\rangle$,
that determines the deviation of probability distribution from
the Poissonian \cite{1,2}. Substituting expression
$a\rightarrow z +\delta\alpha$ into the expressions for
$\langle n\rangle=N\langle a^{\dag} a\rangle$ and
$\langle n^2\rangle=N^2\langle a^{\dag} a a^{\dag} a\rangle=
N^2\langle a^{\dag 2} a^2\rangle+\langle n\rangle$
and after the Talor expansions in the first order of $1/N$,
we have
\begin{equation} \label{12}
F=2 B+\left(\frac{z^*}{z} C+ {\rm c.c.}\right).\end{equation}
We see, that in order to determine the time dependence of the principal
squeezing $S$ in (\ref{11}) and the Fano factor (\ref{12}) for nonlinear
oscillators, we need to find the time dependence of $z$, $C$, and $B$ from
(\ref{9}) for the Hamiltonian (\ref{1}). Following the general procedure
described
in previous section, we first neglect quantum correction $Q/N$ in Eq.
(\ref{9a}). In this case, equation (\ref{9a}) has an exact solution in
the form
\begin{equation}\label{12'}
z(t)=z_0\exp\left[(-i\Delta -\gamma/2) t\right]
\exp\left[ -i g_l |z_0|^{2 l}\mu_l(t)\right],\quad
\mu_l(t)\equiv\left[ 1-\exp( -\gamma l t ) \right]/\gamma l.\end{equation}
We find the differentials $dz$ and $dz^*$
of classical solution (\ref{12'}), and using the substitution
$|dz|^2+\tilde{B}\rightarrow B$ and $(dz)^2\rightarrow C$, we get
\begin{eqnarray}\label{B_C_explicit}
C(t) &=&
-l z_0^2 |z_0|^{2(l-1)}g_l\mu_l(t)\left( l |z_0|^{2 l} g_l\mu_l(t) +i\right)
\exp\left[ (-\gamma -i 2\Delta) t -i 2 |z_0|^{2 l} g_l\mu_l(t) \right],
\nonumber\\
B(t) &=& \exp(-\gamma t)\left[ 1/2+l^2 |z_0|^{4l} g_l^2\mu_l^2(t) \right] +
\left( \langle n_d\rangle + 1/2\right) \gamma t,\end{eqnarray}
where we have taken into account the initial conditions for $B$ and $C$,
Eq. (\ref{initial_cond}).
By substituting formulas (\ref{B_C_explicit}) into Eq. (\ref{11}), we obtain
in the limits
$\tau\equiv g_l(N) t\ll 1$ and $\gamma t\ll 1$ a very simple dependence of $S$
on time as
\begin{equation}\label{14}
S(t)=1-\left[ l x_0^{2l}-(\gamma/g_l)\langle n_d\rangle\right] 2\tau<1,
\end{equation}
where for the sake of simplicity we have assumed that the initial value $z_0$
is real, $x_0={\rm Re} z_0$, and we have taken into account only terms that
are linear in $\tau$ and $\gamma t$. The short-time approximation $\tau\ll 1$
as well as the limit of a large photon number $N\gg 1$ are quite realistic
for a
nonlinear medium modelled by the anharmonic oscillators
(for numerical estimates, see \cite{1}, chap. 10, and \cite{4}). It should
be noted that our formula (\ref{14}) coincides with the corresponding
formula for $S(t)$ in \cite{4} for the Kerr nonlinearity ($l=1$) without
loss ($\gamma=0$). In the case of no loss ($\gamma=0$), our formula
(\ref{14}) shows that the rate of squeezing is determined by the factor
$2 l x_0^{2 l}\lambda_l N^l\equiv 2 l {\cal P}^{(2 l+1)}$. Since
$\lambda_l$ is proportional to the $(2 l+1)$-order nonlinear
susceptibility, the factor ${\cal P}^{(2 l+1)}$ has a physical meaning
of nonlinear polarization. Therefore, the stronger is nonlinear the
polarization induced by light in the medium, the more effective
squeezing of light is possible. For a finite dissipation $\gamma\not=0$,
the squeezing is determined by an interplay between the polarization of
nonlinear medium modelled by the anharmonic oscillator and the thermal
fluctuations of the reservoir. As follows from (\ref{14}), there exists the
a critical number of phonons
$\langle n_d\rangle^{(cr)}=(l/\gamma) {\cal P}^{(2 l+1)}$
such that for $\langle n_d\rangle \ge \langle n_d\rangle^{(cr)}$ the
squeezing is no longer possible.
\par
In the same approximation,
we obtain from (\ref{12}) the following time dependence of the Fano factor
\begin{equation}\label{15}
F(t)=1+2\langle n_d\rangle\gamma t.\end{equation}
Thus, the statistic is super-Poissonian for any $\gamma\not=0$ and is
independent of the degree of nonlinearity $l$.
This is in a good agreement with the earliest result of \cite{2} for
the case of a dissipative Kerr oscillator ($l=1$), where
the impossibility of sub-Poissonian statistics and antibunching
were found from the exact solution.
\par
Turn now to the discussion of the ranges for validity of our approach.
It is easy to see that in terms of our approach the time dependence of
the number of quanta for $l=1$ is
\begin{equation}\label{16}
\langle n\rangle(t)+1/2=
N |z|^2+B
\approx N|z_0|^2 ( 1-\gamma t)+\langle n_d\rangle \gamma t,
\quad \gamma t\ll 1,\quad g_l t\ll 1,\end{equation}
where we have used expressions (\ref{B_C_explicit}) for cumulants $B$
and $C$. It is instructive to compare (\ref{16}) with the exact solution for
$\langle n\rangle(t)$ for the Kerr nonlinearity \cite{2}
\begin{equation}\label{17}
\langle n\rangle(t)=\langle n_0\rangle\exp(-\gamma t)+
[1-\exp(-\gamma t)]\langle n_d\rangle.\end{equation}
Eq. (\ref{16}) and Eq. (\ref{17}) both describe the evolution of an initially
coherent state to a final chaotic state
being characteristic for the reservoir. It is evident that formulas
(\ref{17}) and (\ref{16}) coincide, when $\gamma t\ll 1$ and
$\langle n_0\rangle\simeq N\gg 1$.
A more accurate analysis of the condition for validity of the $1/N$-expansion
should include a comparison of the solution of quantum motion
equation (\ref{9a}), which takes into account the
quantum correction $Q/N$ given by (\ref{10}), with the solution of classical
motion equation (\ref{12'}).
It may be shown after some
algebra, that if $\gamma t\ll 1$ and $\tau\ll 1$,
the influence of the quantum correction $Q/N$ on the dynamics of the mean value
$z$ is of the order $1/N$ and, therefore, our cumulant
expansion is well-defined for $N\gg 1$. The same conclusion could be obtained
considering another criterion of validity (\ref{R-criterion}).

\section{Quantum-classical correspondence in self-pulsing regime of
second-harmonic generation} \label{sec:SHG}

We now consider another example of a quantum optical system, namely
intracavity SHG. The Hamiltonian describing two interacting
quantum modes in the interaction picture has the form
\cite{drummond,savage-pra}
\begin{equation}\label{shg-ham}
H=\sum_{j=1}^2 \Delta_j b_{j}^{\dag} b_{j}+
i E N^{1/2} (b_1^{\dag}-b_1)+
\frac{i\chi}{2} (b_1^{\dag 2} b_2 - b_1^2 b_2^{\dag}),\end{equation}
where the boson operators $b_j$ $(j=1,2)$ describe fundamental and
second-harmonic modes, respectively,
$\Delta_j$ is the cavity detuning of mode $j$, $E N^{1/2}$ is the
classical field driving first mode ($E$ is of order of unity),
$\chi$ is a second-order nonlinear susceptibility. The linear reservoir
and its interaction with a second-order nonlinear medium are described
by the Hamiltonians (\ref{2}).
Now we can rewrite full Hamiltonian of the problem in the form
$H=N {\cal H}$, where ${\cal H}$ has
the same form as (\ref{shg-ham}) and (\ref{2}) with account of the
replacements analogous to (\ref{5}) and definition of new coupling
constant
\begin{equation}\label{shg-g}
g=\chi\sqrt{N},\end{equation}
which is of order unity. Formally, the procedure of the $1/N$-expansion
developed in sec. \ref{sec:1/N-expansion} can not be applied to
the problem of SHG, however
its straightforward generalization to the case of two interacting
modes gives in the first order of $1/N$ the following self-consistent
set of equations
\begin{mathletters} \label{shg-cumulant}
\begin{equation}\label{shg-cumulant_a}
\dot{z}_1=-\frac{\gamma_1}{2} z_1+E + g z^*_1 z_2 +
\frac{1}{N} g B_{12},\end{equation}
\begin{equation}\label{shg-cumulant_b}
\dot{z}_2=-\frac{\gamma_2}{2} z_2 -\frac{g}{2} z^2_1 -
\frac{1}{N} \frac{g}{2} C_1, \end{equation}
\begin{equation}\label{shg-cumulant_c}
\dot{B}_1=-\gamma_1 (B_1-B^{(0)})+g B^{*}_{12} z_1+ g B_{12} z^*_1+
C^{*}_1 z_2 + C_1 z^{*}_2, \end{equation}
\begin{equation}\label{shg-cumulant_d}
\dot{B}_2=-\gamma_2 (B_2-B^{(0)})-g B^{*}_{12} z_1- g B_{12} z^*_1,
\end{equation}
\begin{equation}\label{shg-cumulant_e}
\dot{C}_1=-\gamma_1 C_1+ 2 g ( C_{12} z^*_1+ B_1 z_2),
\end{equation}
\begin{equation}\label{shg-cumulant_f}
\dot{C}_2=-\gamma_2 C_2- 2 g C_{12} z_1,\end{equation}
\begin{equation}\label{shg-cumulant_g}
\dot{C}_{12}=-0.5 (\gamma_1+\gamma_2) C_{12}+ g B_{12} z_2
- C_1 z_1 + C_2 z^*_1, \end{equation}
\begin{equation}\label{shg-cumulant_h}
\dot{B}_{12}=-0.5 (\gamma_1+\gamma_2) B_{12}+ g C_{12} z^*_2
+g z_1 ( B_2-B_1 ),\end{equation}\end{mathletters}
where $z_j\equiv\langle a_j\rangle=N^{1/2}\langle b_j\rangle$,
$B_j=N\langle\delta\alpha^*_j\delta\alpha_j\rangle+0.5$,
$C_j=N\langle\left(\delta\alpha_j\right)^2\rangle$ $(j=1,2)$,
$B_{12}=N\langle\delta\alpha^*_1\delta\alpha_2\rangle$,
$C_{12}=N\langle\delta\alpha_1\delta\alpha_2\rangle$ and
$B^{(0)}$ is defined in Eq. (\ref{B_0}). The
initial conditions for system (\ref{shg-cumulant}) are $B_j(0)=1/2$,
$C_j(0)=C_{12}(0)=B_{12}(0)=0$, $z_2(0)=0$, and $z_1(0)=z_0$,
where $z_0$ is of order unity. In this work, we limit ourselves
only by the values of the field strength $z_0$ corresponding to
self-oscillations \cite{drummond} and $\Delta_1=\Delta_2=0$.
\par
It is easy to see that in the limit $N\rightarrow\infty$ and for
$g={\rm const}\simeq 1$, we get from Eqs. (\ref{shg-cumulant_a}) and
(\ref{shg-cumulant_b}) correct classical motion equations for the
scaled field amplitudes. The solution of motion equations
(\ref{shg-cumulant_c})-(\ref{shg-cumulant_h})
for the second-order cumulants has the form
\begin{equation}\label{shg-solution}
{\bf X}(t)=\overline{\bf X}(t)+
\left( \gamma B^{(0)} t, \gamma B^{(0)} t, 0, 0, 0, 0\right),\quad
{\bf X}(t)\equiv\left[B_1(t),B_2(t),C_1(t),C_2(t),B_{12}(t),C_{12}(t)\right],
\end{equation}
where vector $\overline{\bf X}$ describes the part of {\bf X} that
can be obtained by linearization around a classical trajectory. Variations
near a stable limit cycle rapidly approach zero and therefore
$\overline{\bf X}(t)\rightarrow 0$. As a result, we have only a diffusive
growth of cumulants $B_j$ $(j=1,2)$ as
\begin{equation}\label{shg-dif} B_j(t)=0.5 \gamma_j t,\end{equation}
where we considered the case of a quiet reservoir $\langle n_d\rangle$.
This result indicates that the influence of reservoir zero-point
energy on the dynamics of the nonlinear system is principal physical
mechanism responsible for the difference between the classical and
quantum dynamics in the semiclassical
limit. A time scale $t^*$ for a correct description of the dynamics of
quantized SHG in terms of classical electrodynamics can be found by using
criterion (\ref{R-criterion}). Taking into account that $|z(t)|\simeq 1$,
we have $t^*\simeq 2 N \gamma^{-1}$.
\par
Note that the quantum corrections to the classical motion equations
(\ref{shg-cumulant_a}) and (\ref{shg-cumulant_b})) do not include
cumulants $B_{1,2}$. Therefore,
in the first order of $1/N$, there is no difference between the evolution
of quantum mean values and the classical dynamics for limit cycle. In other
words, the quantum correction $Q\rightarrow 0$, and therefore criterion
(\ref{Q-criterion}) of validity
of $1/N$-expansion does not work. In this respect,
the quantized SHG is a somewhat singular problem. In other quantum optical
systems, for instance, for a nonlinear oscillator with $l\ge 1$, typically
both criteria of validity (\ref{R-criterion}) and (\ref{Q-criterion})
give the same result.
\par
Over a decade ago, Savage addressed
the same problem of quantum-classical correspondence at self-oscillations in
SHG numerically
\cite{savage-pra}. He calculated $Q$ distribution function in the Gaussian
approximation centered on a deterministic trajectory corresponding to a limit
cycle. He worked at a large field and small nonlinearity limits,
$\chi/\gamma_{1,2}\rightarrow 0$, which correspond  to the classical limit
\cite{savage-pra}. It is easy to see that the condition $\chi/\gamma_{1,2}
\rightarrow 0$ is consistent with our condition $N\gg 1$, if one additionally
considers the natural condition of a not very strong dissipation in Eqs.
(\ref{shg-cumulant}), $\gamma_{1,2}/g\lesssim 1$ together with $g\simeq 1$
[Eq. (\ref{shg-g})]. In other words, Savage's small parameter $\chi/\gamma$
corresponds to our large parameter $N$ as $\chi/\gamma\rightarrow N^{-1/2}$.
To establish the difference between the classical and quantum dynamics, the
motion equations for law-order cumulants were obtained in \cite{savage-pra}
and solved numerically for particular values of the parameters \cite{note}.
Based on the results of numerical simulations, Savage concluded that it was
a quantum diffusion that was mostly responsible for the difference between the
classical and quantum dynamics in the semiclassical limit. Moreover, his
numerical estimate for a characteristic time for the classical description
scales as $(\gamma/\chi)^2$, which is in a good
agreement with our analytical result $t^*=2\gamma^{-1}N$. In summary, our
analytical results for the quantum-classical correspondence at self-pulsing
in SHG are quite consistent with the previous numerical investigation of same
problem in \cite{savage-pra}.

\section{Conclusion}
\label{sec:Concl}

We developed the method of $1/N$-expansion to consider the nonlinear
dynamics and nonclassical properties of light in dissipative optical
systems in the limit of a large number of photons. The method was
applied to the
investigation of squeezing in higher-order dissipative nonlinear
oscillators. We would like to note that our method can also be directly applied
to an important case of nonclassical states generation in a medium involving
competing nonlinearities \cite{11}.
\par
We found a time scale of validity of the $1/N$-expansion for a classical
description of the dynamics of nonlinear optical systems with a simple
attractor and a limit cycle. For systems with a simple attractor, this time
scale is of order unity, and for a limit cycle -- proportional to a large $N$.
Qualitatively this result can be understood as follows. For time of order
unity, the trajectory spirals around a stable stationary point with a small
amplitude and therefore in virtue of the uncertainty principle, the
contribution of quantum corrections to the classical motion equations becomes
very important. Unlike the previous case, the
oscillations corresponding to a limit cycle are often close to harmonic and
thus their quantum and classical descriptions can coincide for a fairly long
period of time. The basic difference between the classical
and quantum dynamics in the latter case originates from the influence of
reservoir zero-point fluctuations, which in our notations are of order of $1/N$.
This result is in a good agreement with the result of earliest numerical
simulations of self-oscillations in the quantized second harmonic generation
\cite{savage-pra}.
Finally, it should be noted that our findings are of a rather general nature
and can be applied
to the investigations of self-oscillations in other optical systems,
for example, in the optical bistability \cite{ikeda,lugiato,orozco}.

\acknowledgements

We would like to thank Antoine Heidmann, Evgeny Bulgakov and Zdenek
Hradil for useful discussions.
The work was partially supported by  Czech Grant Agency (grant 202/96/0421)
and Czech Ministry of Education (grant VS96028).


\begin{references}

\bibitem[*]{email1} E-mail:  kna@tnp.krascience.rssi.ru
\bibitem[**]{email2} E-mail: perina@optnw.upol.cz

\bibitem{1} J. Pe\v{r}ina, {\it Quantum Statistics of Linear and Nonlinear
     Optical Phenomena} ( Kluwer Academic Publishers, Dordrecht, 1991).

\bibitem{Fabre}
C. Fabre, Phys. Rep. {\bf 219}, 215 (1992).

\bibitem{Reynaud}
S. Reynaud, A. Heidmann, E. Giacobino and C. Fabre,
         in {\it Progress in Optics XXX},
       edited by E. Wolf (Elsevier, Amsterdam, 1992), p. 1.

\bibitem{7}
A. Heidmann, J. M. Raimond, S. Reynaud, and N. Zagury,
Opt. Commun. {\bf 54}, 189 (1985).

\bibitem{cumulant} L. G. Yaffe, Rev. Mod. Phys. {\bf 54}, 407 (1982).

\bibitem{8} K. N. Alekseev, Opt. Commun. {\bf 116}, 468 (1995).

\bibitem{9} K. N. Alekseev and J. Pe\v{r}ina, Phys. Lett. A {\bf 231},
373 (1997); Phys. Rev. E {\bf 57}, 4023 (1998).

\bibitem{9r}
K. N. Alekseev and D.S. Priimak, Zh. Eksp. Teor. Fiz.
{\bf 113}, 111 (1998) [JETP {\bf 86}, 61 (1998)].

\bibitem{2} V. Pe\v{r}inov\'{a} and A. Luk\v{s}, in {\it Progress in
Optics XXXIII}, edited by E. Wolf (Elsevier, Amsterdam, 1994), p. 130.

\bibitem{4} R. Tana\'{s}, A. Miranowicz, and S. Kielich, Phys. Rev. A
{\bf 43}, 4014 (1991).

\bibitem{drummond}
P. D. Drummond, K. J. McNeil, and D. F. Walls, Opt. Acta {\bf 27}, 321 (1980).

\bibitem{kimble-exp}
H. J. Kimble and J. L. Hall, in {\it Quantum Optics IV}, edited by J. D.
Harvey and D. F. Walls (Springer, Berlin, 1986).

\bibitem{perina-cz-j}
J. Pe\v{r}ina, J. K\v{r}epelka, R. Hor\'{a}k, Z. Hradil, and J. Bajer,
Czechoslovak Journal of Physics B {\bf 37}, 1161 (1987).

\bibitem{savage-pra}
C. M. Savage, Phys. Rev. A {\bf 37}, 158 (1988).

\bibitem{schack}
R. Schack and A. Schenzle, Phys. Rev. A {\bf 41}, 3847 (1990).

\bibitem{szlachetka}
P. Szlachetka, K. Grygiel, J. Bajer, and
   J. Pe\v{r}ina, Phys. Rev. A {\bf 46}, 7311 (1992);
K. Grygiel and P. Szlachetka, Phys. Rev. E {\bf 51}, 36 (1995).

\bibitem{savage-comp}
C. M. Savage, Computers in Physics  {\bf 6}, 513 (1992);
X. Zheng and C. M. Savage, Phys. Rev. A {\bf 51}, 792 (1995).

\bibitem{10}
K. N. Alekseev, G. P. Berman, A. V. Butenko, A. K. Popov,
V. M. Shalaev, and V. Z. Yakhnin,
J. Mod. Opt. {\bf 37}, 41 (1990); Kvant. Electr. {\bf 17}, 425 (1990)
[Sov. J. Quant. Electron., {\bf20}, 359 (1990)].

\bibitem{preprint}
Kirill N. Alekseev and Jan Perina, Physica Scripta, 2000, in press;
Preprint quant-ph/9812019 .

\bibitem{milonni_book}
P. W. Milonni, {\it The Quantum Vacuum: An Introduction to Quantum
Electrodynamics} (Academic Press, Boston, 1994), sec. 2.13 .


\bibitem{note}
The analogous motion equations for cumulants in the problem of SHG were also
considered in the papers \cite{perina-cz-j,schack,szlachetka}. However,
the limit of large photon number was not analyzed in these works.

\bibitem{11}
P. Tombesi, Phys. Rev. A {\bf 39}, 4288 (1989);
M. A. M. Marte, J. Opt. Soc. Am. B {\bf 12}, 2296 (1995); J. Pe\v{r}ina and
J. Bajer, J. Mod. Opt. {\bf 42}, 2071 (1995); C. Cabrillo, J. L. Rold\'{a}n
and P.Garc\'{\i}a-Fernandez, Phys. Rev. A {\bf 56}, 5131 (1997).

\bibitem{ikeda}
K. Ikeda and O. Akimoto, Phys. Rev. Lett. {\bf 48}, 617 (1982).

\bibitem{lugiato}
L. A. Lugiato, L. M. Narducci, D. K. Bandy, and C. A. Pennise,
Opt. Commun. {\bf 4}, 281 (1982); L. A. Lugiato, R. J. Horowicz,
G. Strini, and L. M. Narducci, Phys. Rev. A {\bf 30}, 1366 (1984).

\bibitem{orozco}
L. A.  Orozco, A. T. Rosenberger, and H. J. Kimble,
Phys. Rev. Lett. {\bf 53}, 2547 (1984).

\end{references}
\end{document}